# Photophoretic Light-flyers with Germanium Coatings as Selective Absorbers

*Zhipeng Lu, Gulzhan Aldan, Danielle Levin, Matthew F. Campbell, and Igor Bargatin*

**Abstract:** The goal of ultrathin lightweight photophoretic flyers, or light-flyers for short, is to levitate continuously in Earth's upper atmosphere using only sunlight for propulsive power. We previously reported light-flyers that levitated by utilizing differences in thermal accommodation coefficient (TAC) between the top and bottom of a thin film, made possible by coating their lower surfaces with carbon nanotubes (CNTs). Such designs, though successful, were limited due to their high thermal emissivity (>0.5), which prevented them from achieving high temperatures and resulted in their transferring relatively low amounts of momentum to the surrounding gas. To address this issue, we have developed light-flyers with undoped germanium layers that selectively absorb nearly 80% of visible light but are mostly transparent in the thermal infrared, with an average thermal emissivity of <0.1. Our experiments show that germanium-coated light-flyers could levitate at up to 43% lower light irradiances than mylar-CNT disks with identical sizes. In addition, we simulated our experiments using a combined first-principles-empirical model, allowing us to predict that our 2-cm-diameter disk-shaped germanium-coated light-flyers can levitate in the mesosphere (altitudes 68-78 km) under the natural sunlight (1.36 kW/m$^2$). Similar ultrathin selective-absorber coatings can also be applied to three-dimensional light-flyers shaped like solar balloons, allowing them to carry significant payloads and thereby revolutionize long-term atmospheric exploration of Earth or Mars.

Unmanned aerial vehicles (UAVs) can potentially carry out extensive, long-lasting, and cost-effective exploration of Earth's atmosphere. One proposed kind of small UAVs—sunlight-powered photophoretic flyers, or light-flyers—can levitate in mid-air using a temperature gradient across a nano-cardboard plate [1][2] or by taking advantage of a difference in the thermal accommodation coefficient (TAC) between two sides of a thin film [3].

Previously reported light-flyers were not yet suitable for deployment in Earth's upper atmosphere or on Mars because they required irradiance levels 4-8 times that of natural sunlight, i.e., 1.36 kW/m$^2$ under normal incidence in the mesosphere or stratosphere, approximately 1 kW/m$^2$ at sea level on Earth, and <0.6 kW/m$^2$ on Mars. For example, we previously reported levitation of 6-mm-diameter mylar-CNT disks at irradiances above 5 kW/m$^2$ [3] and 8-mm-wide nano-cardboard plates at irradiances above 10 kW/m$^2$ [1], with both done at room temperature in a vacuum chamber.

Recent theoretical studies suggest the possibility of not just levitation, but also of carrying significant payloads, in Earth's upper atmosphere, where the ambient temperatures are much lower than at sea level. For instance, Schafer et al. [2] predicted that a 10-cm-diameter disk using alumina-based nano-cardboard composites could carry a 0.3-gram payload in the stratosphere, roughly 10 times the craft's mass and sufficient for simple silicon-based sensors. Additionally, Celenza et al. [4] simulated three-dimensional meter-scale porous structures capable of carrying much larger kilogram-scale payloads in the upper mesosphere (see Fig. 1(b)). Though holding significant promise, these results have yet to be practically demonstrated. Laboratory experiments are therefore crucial to confirm the lift forces generated by thermal transpiration or Knudsen pump mechanisms and to move toward real-world applications of these light-flyers.

All other things being equal, the photophoretic force increases with decreasing thermal emissivity, because this causes the light-flyer to be hotter relative to the ambient atmosphere, or increasing TAC difference between the two surfaces, because this results in a greater momentum differential between gas molecules colliding with the top and bottom surfaces [3]. Since all TAC values in air are fairly close to unity due to a molecularly thin organic layer spontaneously formed under ambient conditions [5], it is difficult to dramatically increase the difference in TAC by 10% even for well-characterized metal surfaces [6][7]. However, it is possible to create light-flyers with very low thermal emissivity values by coating them with a selective absorber that has a high absorptivity for the solar spectrum but low emissivity in the thermal infrared (5-15 µm wavelength). Several indirect-bandgap semiconductors, such as germanium and gallium arsenide, possess such optical properties because their bandgaps fall at energies between those of the visible and infrared spectra (i.e., 0.7~1.4 eV [8]). Among these options, undoped germanium is nearly transparent in the thermal infrared spectrum [9] and thus has very low emissivity, making it an ideal candidate for our selective absorber.

As described below, our use of germanium resulted in a significant reduction in the light irradiance needed for levitation from 2.5 kW/m$^2$ CNT-coated light-flyers to just 1.5 kW/m$^2$ for germanium-coated flyers. To corroborate our findings, we also measured the optical and surface properties of germanium and developed a theoretical model to predict the light-flyer's levitating performance in both the laboratory and in Earth's mesosphere. Importantly, our model predicts that natural sunlight would be sufficient for levitation in the mesosphere because the ambient temperature there are much lower than in our laboratory experiments.

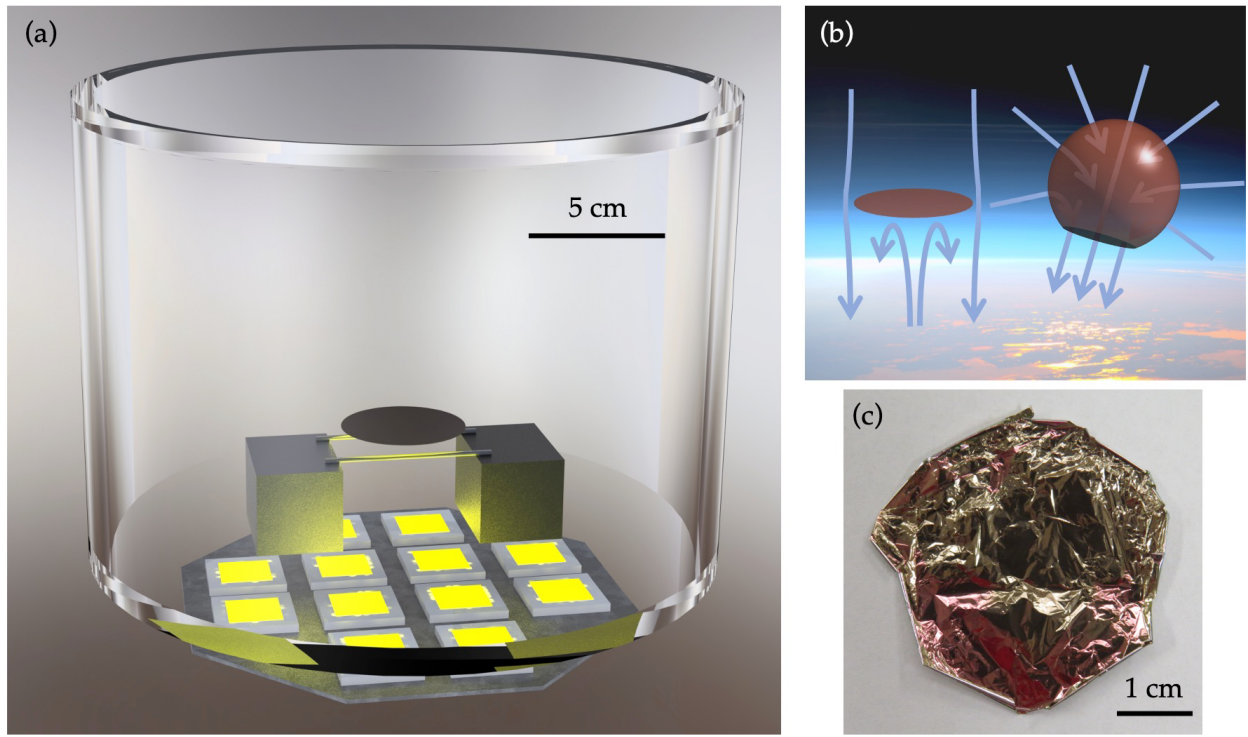

FIG. 1. (a) Schematic diagram of the experimental setup consisting of an acrylic vacuum chamber, a 4-cm-diameter germanium-coated light-flyer, two 2.4-mm-diameter aluminum rods on two 4.5-

cm-tall substrates, and 12 LEDs below the chamber. (b) Conceptual diagram of a germanium-coated light-flyer disk (left) and solar-photophoretic balloon (right) in the mesosphere. The background is courtesy of NASA [10]. (c) Photograph of a 4-cm-diameter alumina-germanium-mylar-germanium light-flyer.

We fabricated disk-shaped light-flyers by sputtering germanium as the light-absorbing layer on both sides of a 500-nm-thick mylar film, followed by atomic-layer-depositing (ALD) a 50-nm-thick layer of alumina onto one side of the film to enhance its light-flyer's mechanical and thermal robustness (see *Supplementary Materials* for detail). Depositing alumina on both sides creates a symmetric structure resulting in no significant TAC difference between the top and bottom sides. Hence, for the purpose of this study, only one side of the film has the alumina. In the final stage of the fabrication process, we laser-cut circular disks of different sizes from our germanium-mylar-germanium-Alumina films.

We tested our microflyers in a cylindrical acrylic vacuum chamber and illuminated them from below using an arrangement of twelve 100-W LED arrays, which created irradiance of up to 6 kW/m$^2$ (6 Suns). In a single experiment, we held the pressure constant while gradually increasing the light irradiance until the light-flyer lifted off; we then repeated this procedure at different pressures.

Vacuum chamber levitation experiments can be impacted by the ground effect, in which an artificial lift force is generated as the light-flyer interacts with the chamber floor [11]. To avoid overestimating the lift force due to this effect, we followed the guidelines from Ref. [11] for minimizing the launchpad-associated ground effect. This involved using a sparse launchpad consisting of two 2.4-mm-diameter aluminum rods, as shown in Fig. 1 (a). These thin aluminum rods exhibited a minimal ground effect similar to that of J-shaped steel wires from Ref. [11], as shown in Fig. S2. We used the rods rather than the wires because they were more convenient in sample loading and testing. We also positioned the launchpad about 4.5 cm above the chamber floor, a distance greater than the diameter of the largest sample, in order to minimize the floor-associated ground effect [11].

Figure 2 illustrates the impact of the selective-absorption coating by comparing the levitation of germanium-coated light-flyers to those with previously reported CNT-coatings [3][11]. It can be seen that, for all disk sizes, the irradiance required for levitation is lower for Ge-coated light-flyers, demonstrating the impact of this type of coating. Specifically, the minimum irradiances needed for levitation were 29%, 39%, and 43% lower for 2, 3, and 4-cm-diameter germanium light-flyers, respectively. The significant decrease can be attributed to much lower emissivity of germanium, which increases the temperature of germanium-coated samples compared to CNT-coated ones. Specifically, for the 2-cm-diameter disks, our model (discussed later) predicts a 441 K temperature for the germanium-coated samples at its optimal pressure, whereas that for the CNT-coated samples is 420 K. Larger disks exhibited greater reductions in minimum irradiances due to their higher surface area and thus higher importance of radiative heat dissipation and lower relative importance of air conduction. Mathematical details of the emissivity's effect on the disk temperature and photophoretic force are described in the *Supplementary Materials*.

Plotting the minimum irradiance needed for levitation versus the chamber pressure shows a parabola-like minimum in the log-log scale (Fig. 2). The photophoretic force was both predicted [12][13][14] and observed [11] to maximize at pressures where the mean free path (MFP) is similar in magnitude to the disk diameter (i.e., for 0.01 < Kn < 10, where Kn is the Knudsen number, equal to the ratio of the gas's mean free path to the diameter of the levitating disk). The bottom of the parabola corresponds to the *optimal pressure* that maximizes the lift force.

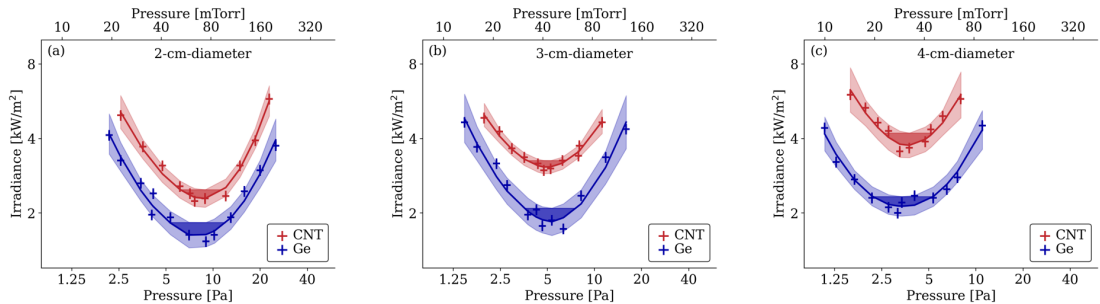

FIG. 2. Levitation performance of (a) 2, (b) 3, (c) 4-cm-diameter alumina-mylar disks with Ge or CNT coatings. The flyers were launched from two aluminum rods (2.4 mm in diameter) that were placed far from the vacuum chamber's bottom (4.5 cm). Solid lines show locally estimated scatterplot smoothing (LOESS) fits, with 99% confidence intervals marked with light shading. Estimated optimal pressures and minimum irradiances are shown with darker shading.

We extensively characterized our films in order to deepen our understanding of their optical and TAC properties. Our experiments included measuring the optical properties of our films using both thermal imaging and transmittance-only Fourier-transform infrared (FTIR) spectrometry aimed at testing our hypothesis that germanium leads to a lower emissivity than CNTs. We also measured the films' TAC values [6] and their roughness using atomic force microscopy (AFM).

The ability of materials to lose heat by radiation is characterized by their emissivity ($\varepsilon$), which is related to their transmittance ($t$) and reflectance ($r$) at the same wavelength through the equation $\varepsilon = 1 - t - r$. First, we estimated the transmittance and reflectance at thermal infrared wavelengths using thermal imaging. To conduct this test, as shown in Fig. 3 (a, b), we used a hotplate as a radiation source and measured thermal infrared radiation transmitted and reflected by a light-flyer film at normal incidence using a thermal camera. The details are described in the *Supplementary Materials*. Subtracting 90-degree transmittance and reflectance from unity resulted in the emissivity of approximately 0.08 and 0.48 for germanium and CNT-based films, respectively, as tabulated in Table 1. Using a similar measurement in the optical frequency range, we also estimated the visible-range absorptivity of both CNT and germanium to be 0.75 using a white flashlight as the light source.

TABLE 1. Optical properties and TAC for various materials.

| **Property** | **Mylar** | **Mylar-Ge** | **Mylar-CNT** |
|---|---|---|---|
| IR transmittance | 0.92±0.01 | 0.81±0.01 | 0.47±0.02 |
| IR reflectance | 0.02±0.01 | 0.11±0.01 | 0.05±0.01 |
| IR emissivity | 0.06±0.02 | 0.08±0.02 | 0.48±0.03 |
| Visible transmittance | 0.94±0.01 | 0.18±0.01 | 0.22±0.01 |
| Visible reflectance | 0.02±0.01 | 0.07±0.02 | 0.03±0.01 |
| Visible absorptivity | 0.03±0.02 | 0.75±0.03 | 0.75±0.02 |
| TAC | 0.77±0.03 | 0.90±0.03 | 1.08±0.05 |

As shown in Fig. 3 (c), we also measured the transmittance of mylar, mylar-CNT, and mylar-germanium light-flyer materials at wavenumbers from 4000 to 400 cm$^{-1}$ using a FTIR spectrometer. To estimate an upper bound for emissivity, we assumed zero reflectance of the films, giving the upper bound $\varepsilon_{max} = 1 - t$. Based on the expected temperatures of 400 to 500 K for a light-flyer under sunlight, spectral intensity reaches a maximum at approximately 900 cm$^{-1}$, corresponding to 11 $\mu$m wavelength. Therefore, if we averaged the transmittance in the range 900±400 cm$^{-1}$, where the blackbody radiation peaks at temperatures between 400-500 K, we could derive the upper bound for emissivity to be 0.11 and 0.49 for germanium and CNT-based films, respectively, values that agree with those obtained through thermal imaging. Our results show that germanium has nearly five times lower emissivity than CNTs, resulting in significantly larger temperatures and a stronger photophoretic lift force.

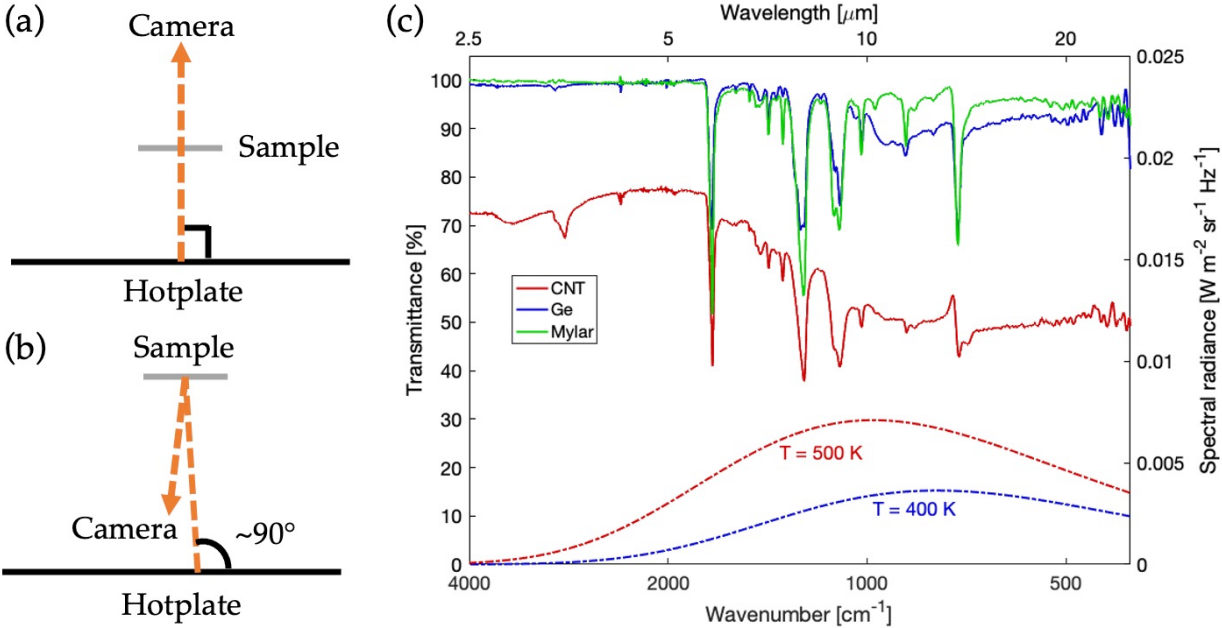

FIG. 3. Characterization results of optical properties from (a, b) thermal imaging and (c) transmittance-only FTIR. Schematics of the thermal imaging method measuring the 90-degree (a) transmittance and (b) reflectance. (c) The left axis plots the transmittance of various films as characterized by FTIR, while the right plots the calculated spectral radiance of black body radiation at 400 and 500 K surface temperature. The shaded area highlights the spectrum where transmittance is averaged.

Next, we characterized the TAC values of various surfaces using the experimental setup described in Ref. [6]. We first calibrated our experimental setup by comparing results of gold and platinum with literature values. We then prepared the same samples as for FTIR and averaged three sets of results for each of mylar, mylar-CNT, and mylar-germanium, and mylar-alumina, as tabulated in Table 1. Specifically, mylar-alumina had a TAC of 0.79±0.03. The measured TAC values between 0.8 and 1.1 in air are consisted with literature reports that the TAC is usually close to unity for interactions with relatively heavy molecules, such as nitrogen and oxygen [6][7][15]. We note that, although the TAC is defined to be between 0 and 1, the measured TAC may be slightly larger than unity due to the systematic measurement errors [16][17]. However, the differences between the TACs of any two materials, which is the quantity needed to predict photophoretic forces, are accurate even in the presence of such systematic errors. Our measurements, summarized in Table 1, show that the typical TAC difference between a top layer (e.g., mylar or alumina) and a bottom one (i.e., germanium or CNT) is approximately 0.1 for germanium-based light-flyers and 0.25 for CNT-based ones. Details are described in the *Supplementary Materials*.

Surface roughness is one of the factors affecting TAC, with greater roughness typically resulting in a higher TAC because an incident molecule is more likely to collide multiple times with a rough surface and therefore reach a higher degree of thermalization. To better understand our TAC findings, we conducted Atomic Force Microscopy (AFM) of various surfaces we used for light-flyers. As shown in Fig. 4 (b-d), the root mean squared (RMS) roughness of mylar, mylar-germanium, and mylar-CNT were 2.5, 34.8, 71.2 nm, respectively. This upward trend aligns well with the TAC values presented in Table 1, where smoother surfaces have a lower RMS roughness and consequently lower TAC. Note that the sharp peaks observed in Fig. 4 (a, b) are associated with sub-micron particles intentionally included into the mylar sheet by the manufacturer to reduce stiction. These rare inclusions do not affect the number of collisions between a molecule and the surface and therefore were not included in the RMS roughness comparison. Environmental Scanning Electron Microscopy (e-SEM) images of these particles on both mylar and mylar-germanium surfaces are shown in Fig. S4.

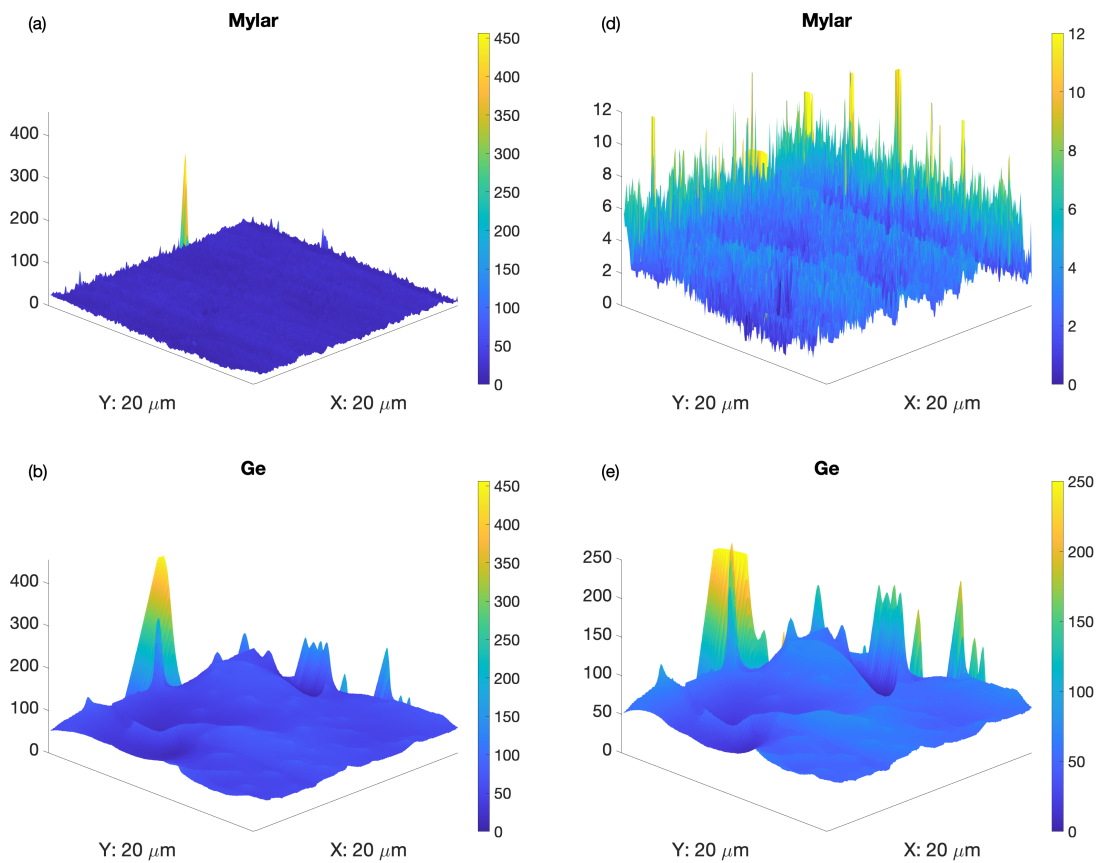

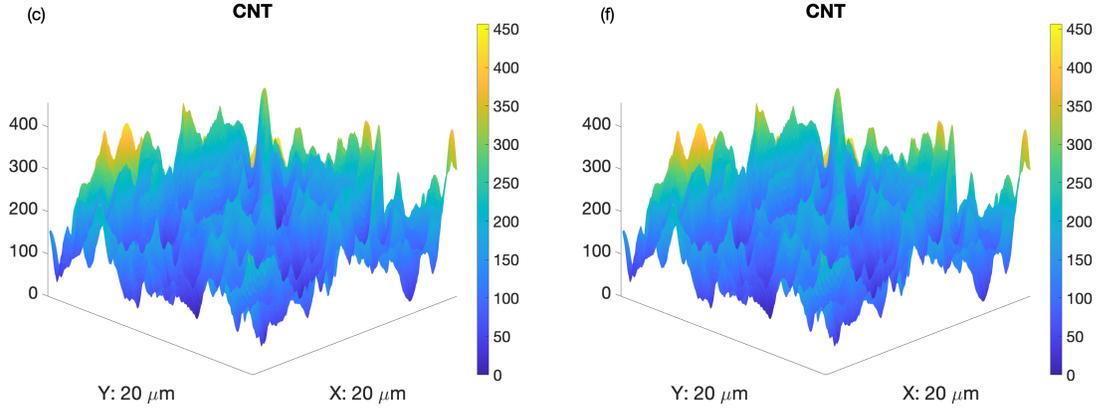

FIG. 4. Characterization results of surface properties from AFM. (a-c) are three-dimensional surface roughness graphs focusing on 20 $\mu m$ by 20 $\mu m$ spots for (a) mylar, (b) mylar-Ge, and (c) mylar-CNT surfaces. (d-f) are rescaled graphs to exclude friction enhancement particles.

Despite their lower TAC difference, the much lower emissivity of germanium-coated light-flyers still offers net advantages in real-world levitation applications in Earth's mesosphere or on Mars. It is therefore important to conduct experiments in a low-temperature vacuum chamber at the conditions of the mesosphere. While state-of-the-art techniques can cool down the vacuum chamber by approximately 100 K using dry ice or multiple vacuum cooling methods [20][21], these techniques are expensive and challenging to implement, especially under high optical flux. Instead, we performed experiments at room temperature and used them to develop a revised model for our light-flyers based on the photophoresis theory and our experimental data.

According to Rohatschek's work [13], the semi-empirical equation of the photophoretic force is $1/F = 1/(C_{fm}F_{fm}) + 1/(C_{cont}F_{cont})$, where $F$ stands for the photophoretic force, subscripts $fm$ and $cont$ for free-molecular and continuum regime, respectively, and $C_{fm}$ and $C_{cont}$ are typically assumed one for atmospheric aerosols [13][22]. Equations for the derivation of light-flyer's temperature and photophoretic force are discussed in the *Supplementary Materials*. However, the case of $C_{fm} = C_{cont} = 1$, where spherical microparticles were levitated, did not agree well with our experimental observations of ultrathin disks. As shown in Fig. 5, when $C_{fm} = 0.4$ and $C_{cont} = 3$, the theoretical irradiance-pressure curves agreed the best with our experimental observations for all light-flyer sizes of both germanium and CNT-coated samples at all pressures.

The inclusion of $C_{fm}$ and $C_{cont}$ is motivated by the recognition that the original semi-empirical interpolation method is most accurate in either free-molecular or continuum regimes and is least accurate in the transition regime. However, the $\Delta\alpha$-photophoretic force reaches its maximum in the transition regime where the physics involved is complex, which necessitates adjustments to these two coefficients. When dealing with significantly lower or higher pressures, $C_{fm} = C_{cont} = 1$ remains valid. Nevertheless, fitting is necessary in the transitional flow region before a accurate and comprehensive model can be developed specifically for the transition regime, which is most useful in practical applications.

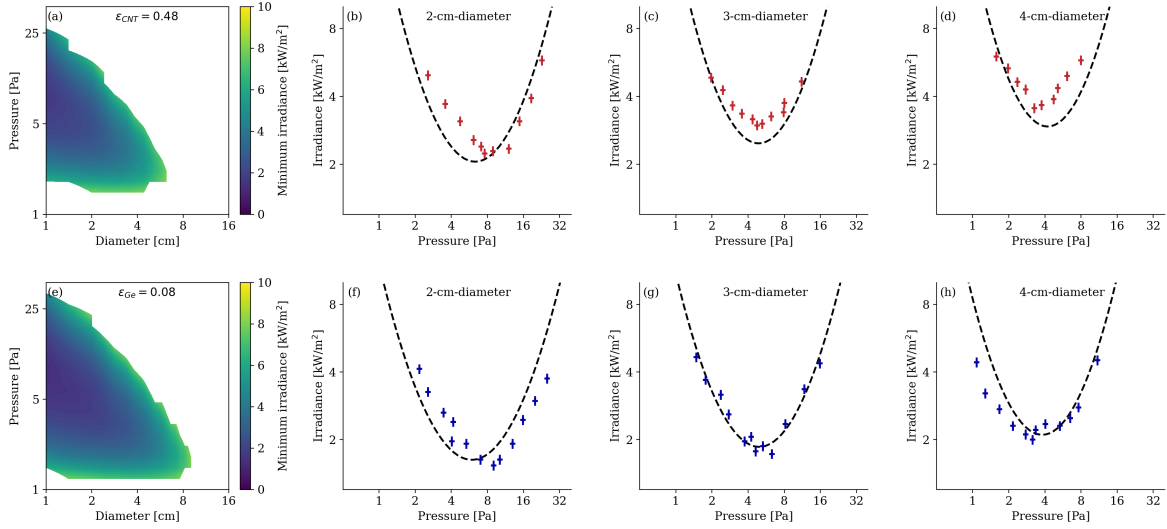

FIG. 5. Theoretical predictions and experimental data. (a) Contour maps of the disk temperature for (i) CNT-coated disks (by assuming a uniform emissivity $\varepsilon_{CNT} = 0.48$) and (ii) germanium-coated disks (by assuming $\varepsilon_{Ge} = 0.08$). (b-d) Predicted irradiance-pressure pairs and the experimental irradiance-pressure data for (i) CNT-coated disks and (ii) germanium-coated disks with (b) 2, (c) 3, and (d) 4 cm diameters.

This theoretical modeling of the light-flyer's performance in our laboratory setting can be applied to Earth's mesosphere as well. Earth's mesosphere extends from 50 to 100 km in altitude [18][19], with pressures ranging from 75 to 0.025 Pa (i.e., 560 to 0.19 mTorr). The ambient temperatures in the mesosphere vary between 180 and 270 K, significantly lower than the room temperature in our vacuum chamber (293 K). The high-altitude temperature profile therefore offers considerable room for performance improvements, such as more effective radiative heat dissipation and less thermal deformation of mylar.

Take the 2-cm-diameter germanium-coated light-flyer as an example because it had a lower minimum irradiance than the 3 and 4-cm-diameter disks. As shown in Fig. 6 (a), assuming a maximum insolation of 1.36 kW/m² on top of Earth's atmosphere [23][24], this light-flyer could levitate between 68 and 78 km altitude, corresponding to a pressure range of 6.3 to 1.3 Pa (i.e., 47 to 10 mTorr). A more conservative estimate can be considered by assuming an incident angle of the sunlight smaller than 90 degrees. Without a light-flyer steering mechanism like attaching a payload to an angled and rigid shaft to adjust the tilt angle [2], even at a 45-degree incident angle, the insolation would still be nearly 1.0 kW/m², narrowing the altitude range to 70~76 km. In Fig. 6 (b), the light-flyer temperature is as low as 300 K, which significantly reduces thermal deformation of mylar-based light-flyers and extends its lifespan. Hence, these results suggest that a long-lasting levitation in a 6-km-wide altitude range may be possible, even without a solar tracking mechanism and despite strong wind shears in the mesosphere [24][25]. The outcomes for 3- and 4-cm-diameter disks are presented in Fig. S7, indicating a gradual reduction in the operational pressure range.

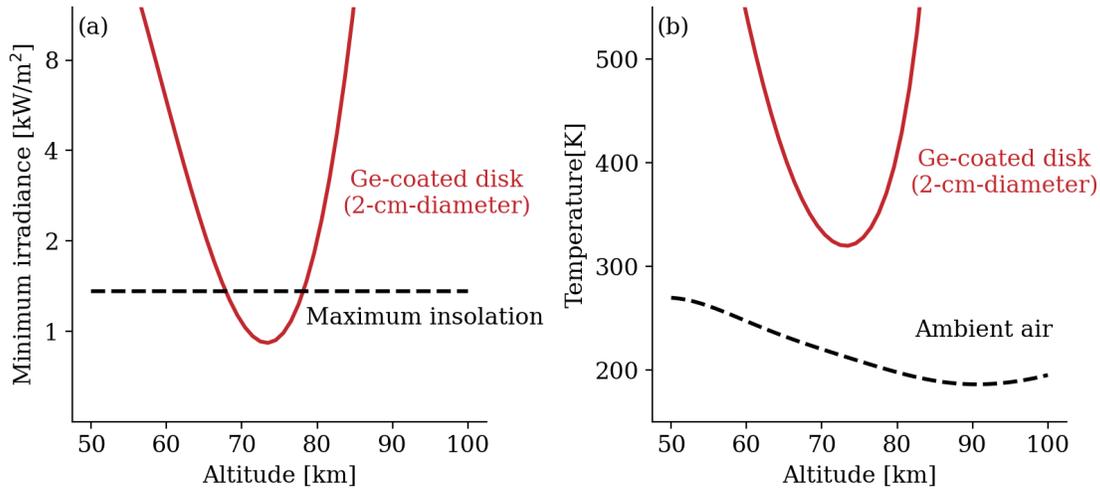

FIG. 6. Key predictions from the theoretical model. All plots assume a germanium-coated light-flyer with top TAC $\alpha_{top} = 0.9$, bottom TAC $\alpha_{bot} = 0.8$, uniform emissivity $\varepsilon = 0.08$, optical absorptivity $\epsilon_{vis} = 0.75$, and areal density $\rho = 1.0$ g/m². (a) The minimum irradiance required for a 2-cm-diameter light-flyer vs. altitude. The black dashed line stands for the maximum insolation (i.e., 1.36 kW/m²). (b) The temperature profiles of this 2-cm-diameter light-flyer and the ambient air versus altitude.

Maximum payloads can also be derived from the model, as discussed in the *Supplementary Materials*. For light-flyer disks a few centimeters in diameter, the payload is generally in the milligram range, which is challenging for practical applications. However, Celenza et al. recently showed that 3D structures made from nanocardboard can increase the payload to the kilogram range [4]. Using germanium as a light-absorbing coating in these structures greatly increase (approximately double) the photophoretic force by reducing the thermal emissivity.

Conclusion

In summary, we have demonstrated light-flyers composed of alumina and mylar as structural layers and germanium as a selective-absorber. We conducted tests on germanium-coated disks with diameters of 2, 3, and 4 cm under minimized ground effect conditions and achieved successful levitations at irradiance levels as low as 1.5 kW/m² at room temperature. We characterized the optical and surface properties of germanium and CNT-based films and found germanium to have a sixfold lower emissivity. Using experimental data and semi-empirical photophoretic force, we developed a revised model and predicted levitation capabilities at various altitudes for light-flyers of multiple sizes.

In addition to the two-dimensional photophoretic light-flyer disks, the germanium coating can be applied on three-dimensional geometries like Celenza's [4] balloon designs to reduce radiative heat loss and increase the photophoretic force, paving the way for future low-cost and clean-energy atmospheric research.

# Supplementary Materials

*Zhipeng Lu, Gulzhan Aldan, Danielle Levin, Matthew F. Campbell, and Igor Bargatin*

*1. Vacuum chamber and light emitting diode (LED) setup*

The vacuum chamber was set up in the same way as described in Lu *et al*'s work [1] and as shown in Fig. S1. We used 12 LEDs (LOHAS LH-XP-100W-6000K) to create a more uniform light source with a maximum light irradiance approximately 7 kW/m$^2$.

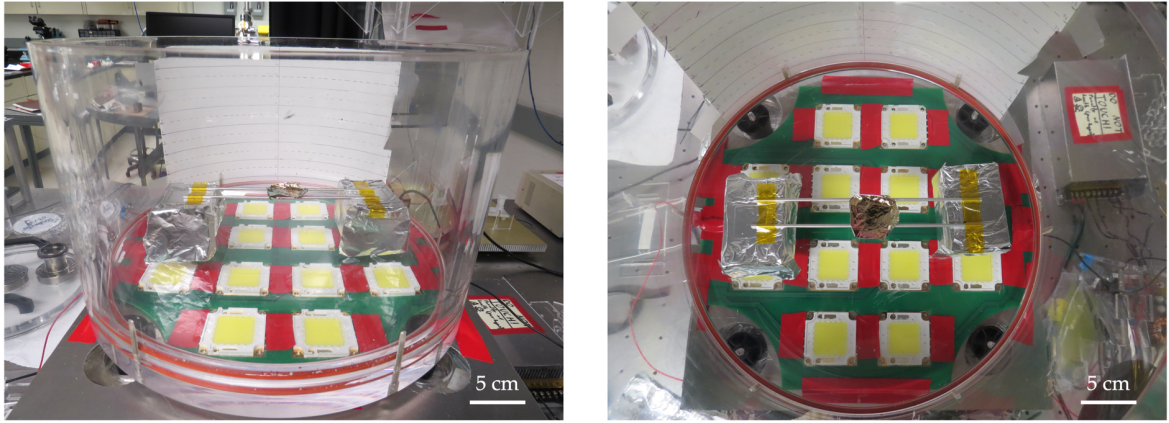

FIG. S1. Photographs of experimental setups consisting of an acrylic vacuum chamber, a 4-cm-diameter germanium-coated light-flyer, two 2.4-mm-diameter aluminum rods on two 4.5-cm-tall substrates, and 12 LEDs below the chamber. (a) Side view. (b) Top-down view.

*2. Launchpad selection*

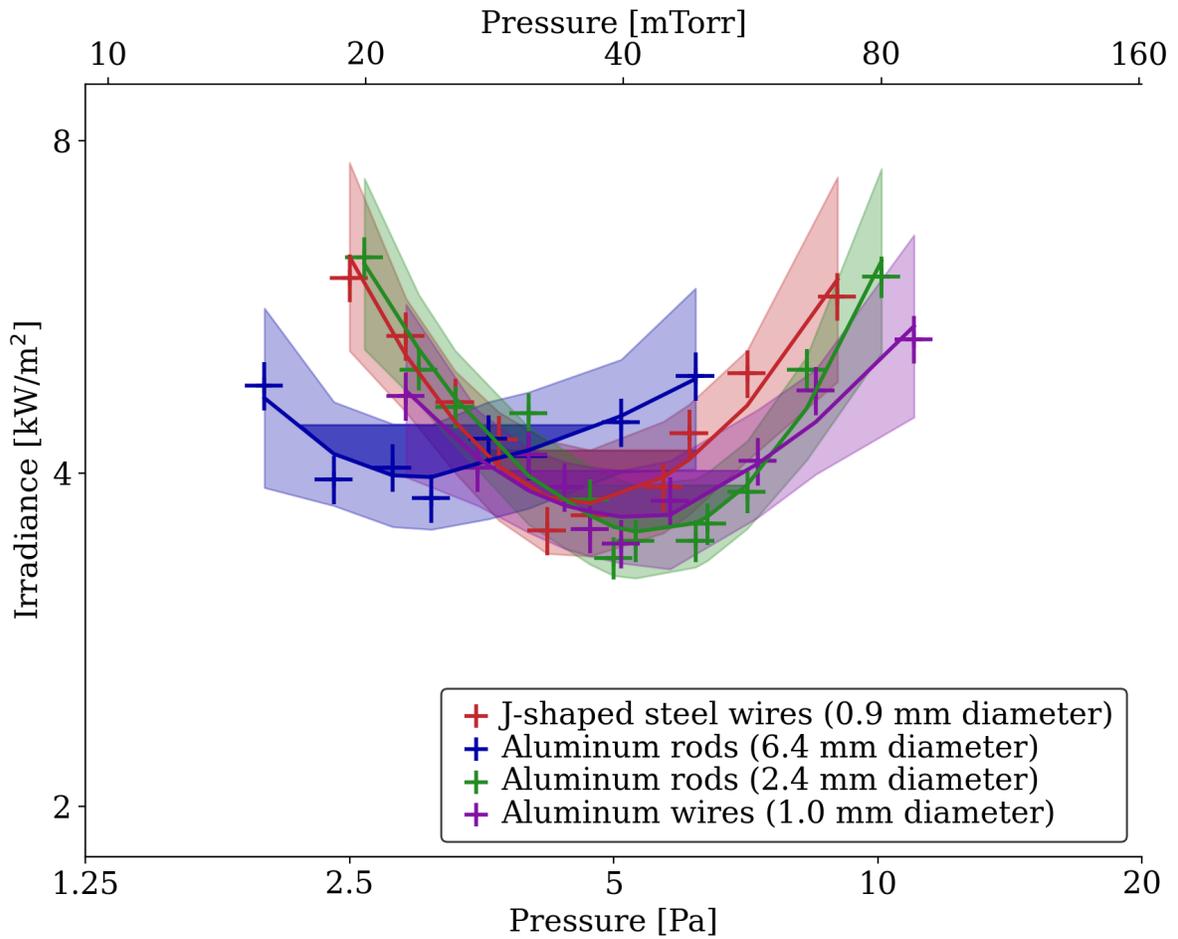

FIG. S2. Same as Fig. 2 for 4-cm-diameter germanium-coated samples but for different launchpads. J-shaped steel wires (0.9 mm diameter), aluminum rods (2.4 mm diameter), and aluminum wires (1.0 mm diameter) exhibited minimal ground effects.

## 3. Levitation performance of another set of light-flyers

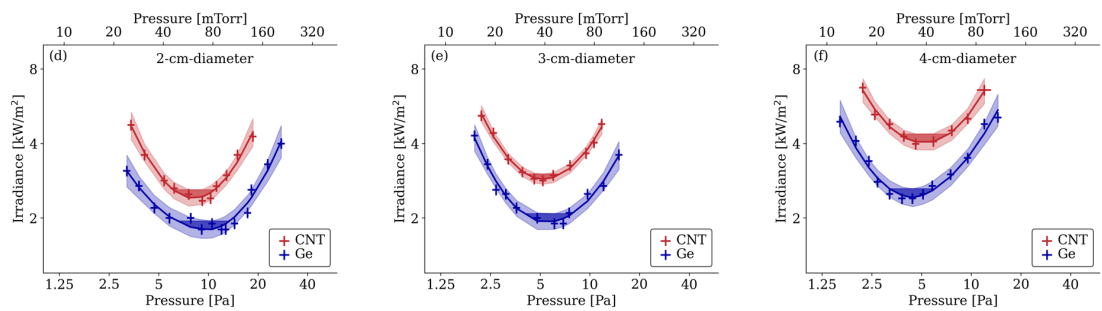

FIG. S3. Same as Fig. 2 but for different CNT and germanium-coated samples with various diameters.

*4. Light-flyer fabrication procedure*

We used a 500-nm-thick mylar sheet (Dupont) as the substrate for all light-flyers. We first wrapped it on a 525-$\mu$m-thick silicon wafer for further steps. Specifically,

(1) *For CNT-coated samples*. We spin-coated an aqueous solution of CNT-COOH water paste (0.2% in weight; NanoAmor) on top of the mylar sheet at 300 revolutions per minute (RPM) for 10 seconds. We then baked the resulting film on a hotplate at 90~100 °C and flipped it when dry. Next, we deposited 50-nm-thick alumina *via* atomic layer deposition (Cambridge Nanotech S200 ALD) at 140 °C using water and $Al_2(CH_3)_6$ precursors. The final film had a multilayer structure of alumina (50 nm)-mylar-CNT (~300 nm).

*(2) For germanium-coated samples*. We deposited a conformal 20-nm-thick layer of germanium *via* physical vapor deposition (Lesker PVD75 DC/RF Sputterer) at 120 W power. We flipped the film and deposited another 20-nm-thick layer of germanium. Likewise, we deposited a 50-nm-thick layer of alumina. The final film had a multilayer structure of alumina (50 nm)-germanium (20 nm)-mylar-germanium (20 nm).

Last, we used laser micromachining (IPG IX-255 Excimer Laser Micromachining) to cut circular disks with 2, 3, and 4 cm diameters. The areal density of each final disk was measured on an analytical balance (AD HR-202), typically ranging from 1.0~1.3 $g/m^2$.

*5. Characterization details*

(1) *For FTIR.* We used the model JASCO FT/IR-6300 with ATR Pro One attachment. We used a scan rate of 16 in the range of 4000-400 cm$^{-1}$ with a 4 cm$^{-1}$ resolution.

(2) *For thermal imaging.* As shown in Fig. 3 (a, b), a hotplate or flashlight was used as a radiation source and placed such that the angle between the incident flux and the surface of the film was approximately 90°. The temperature of the films before being exposed to radiation was the same as the ambient temperature, $T_\infty$ (20°C), which was constant during the measurements. Consider a small part of the hotplate surface that has the same area as the thin film. Given that the distance between the radiation source and the samples was small (i.e., 8~10 cm), we assume that all the flux leaving that specific surface was equal to the incident flux on the film's surface. Three temperatures were recorded by the infrared camera: the temperature of the hotplate, the temperature of the lower side of the film facing the hotplate, and the temperature of the upper side of the film, denoted as $T_{hotplate}, T_{lower},$ and $T_{upper}$, respectively. We assume that $T_{lower}$ mainly results from the 90-degree reflectance with a negligible emissivity due to the film's diffuse radiation. Under above assumptions, we obtain the following equations:

$$r = \frac{T_{lower}^4 - T_\infty^4}{T_{hotplate}^4 - T_\infty^4}$$

$$t = \frac{T_{upper}^4 - T_\infty^4}{T_{hotplate}^4 - T_\infty^4}$$

where $r$ is the reflectance and $t$ is the transmittance of the film.

(3) *For AFM.* We used the model Asylum AFM. The cantilever had a resonance frequency of 310 kHz with drive amplitude 1.38 V.

(4) *For e-SEM.* We used the model FEI Quanta 600 ESEM. The water vapor was kept at 0.38 torr and the voltage was at 15 kV.

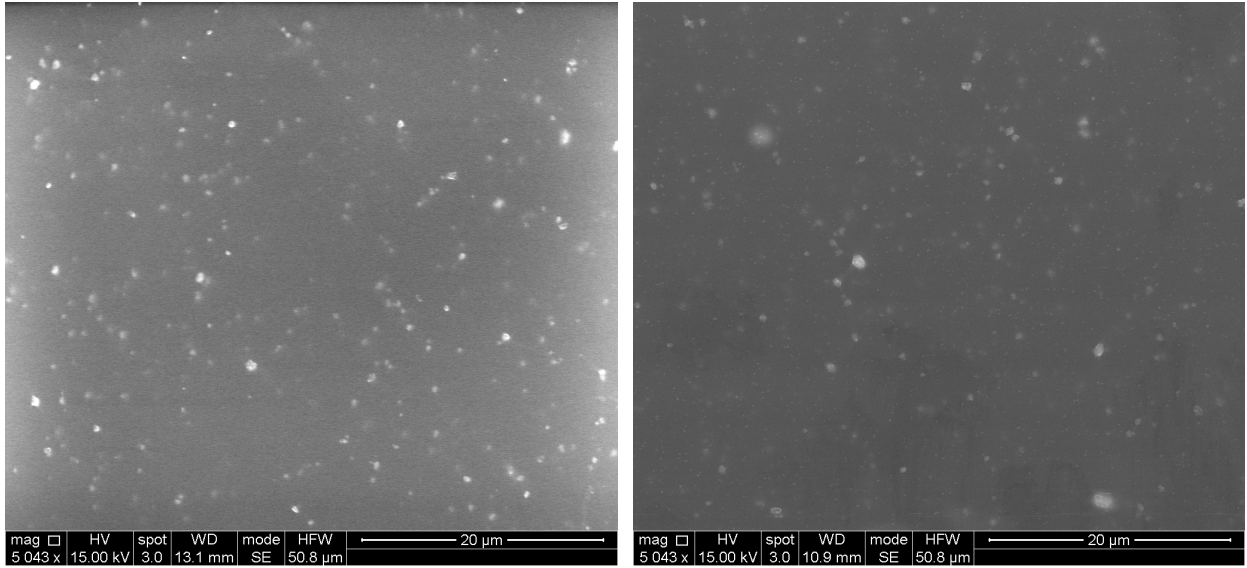

FIG. S4. E-SEM images of (a) mylar and (b) mylar-germanium surfaces.

## 5. TAC measurement and data processing

There are mainly three common geometries for TAC measurements: infinite parallel plates, coaxial cylinders, and concentric spherical shells [2][3]. We built up our measurement system in a larger cylindrical steel chamber to approximate the concentric configuration. In the experimental setup (Fig. S2), a thin circular heater (MCH Ceramic Heating Plate 12V) was hung in the center of the chamber and treated as the internal hot shell, while the whole chamber's wall was treated as the external cold shell. A testing film was affixed on the heater using a tiny amount of silver paste (Arctic Silver 5 Polysynthetic Thermal Compound) and covered over 95% of the heater's surface. It is highlighted that the heater was far from the chamber's wall and that the chamber's surface area was more than two orders of magnitude larger than that of the heater, both of which guarantee a reasonable approximation of our experimental setup to a concentric configuration.

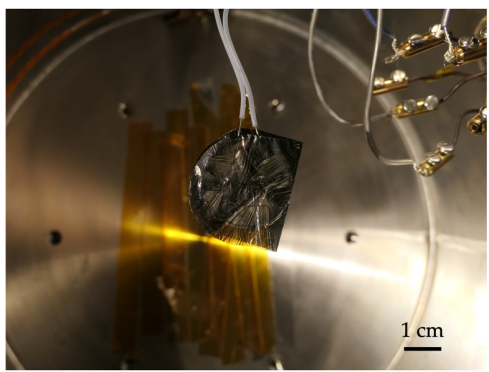

FIG. S5. Photograph of experimental setups consisting of a 24-mm-diameter ceramic heater and a thin mylar-CNT film wrapping the heater inside a cylindrical steel chamber.

TABLE S1. Descriptions of symbols in the TAC measurement methodology.

| SYMBOL | DEFINITION |
|---|---|
| $T_c$ | Cold (ambient) temperature, i.e., 293 K |
| $T_h$ | Hot (heater) temperature |
| $\bar{T}$ | Average temperature, i.e., $(T_h + T_c)/2$ |
| $\Delta T$ | Temperature difference, i.e., $(T_h - T_c)$ |
| $r$ | Radius of the heater |
| $A$ | Surface area of the heater |
| $A_{eff}$ | Effective surface area of the heater |
| $\delta$ | The ratio of the heater's effective surface area to its surface area, i.e., $A_{eff}/A$ |
| $p$ | Ambient pressure |
| $R_{air}$ | Ideal gas constant of air, i.e., 287.05 J/ (kg K) |
| $\bar{v}$ | Average molecular velocity of air |
| $\mu$ | Dynamic viscosity of air |
| $\rho$ | Density of air |
| $D$ | Thermal diffusivity of air |
| $\gamma$ | Adiabatic index of air (approximated as a diatomic gas) |
| $\kappa$ | Thermal conductivity of air |
| $\alpha$ | Thermal accommodation coefficient |
| $g$ | Gravity of Earth, i.e., 9.8 m/s² |
| $Ra$ | Rayleigh number |
| $Nu$ | Nusselt number |
| $\sigma$ | Stefan-Boltzmann constant, i.e., 5.67× $10^{-8}$ W/ (m² K⁴) |
| $\varepsilon$ | Emissivity of disk |
| $U$ | Voltage of the power supply |
| $I$ | Current of the power supply |
| $U_{heater}$ | Voltage of the heater |
| $R_{heater}$ | Resistance of the heater |
| $Q_{in}$ | Electrical input power |
| $Q_{rad}$ | Radiation loss |
| $Q_{cont}$ | Convection and conduction loss in the continuum regime |
| $Q_{fm}$ | Convection and conduction loss in the free-molecular regime |
| $Q_c$ | Convection and conduction loss in all regimes |

The governing equation of the concentric configuration measurement is:
$$Q_{in} = Q_{rad} + Q_c + C\Delta T \tag{1}$$
Where $C$ is a fitting parameter to be fitted together with $\alpha$. The physics behind Eq. (1) is that beside radiation, conduction, and convection loss from air, the heater-sample system also dissipates energy to the environment by conduction loss from the lateral or uncovered part of the heater and Joule's heating. Another check comes from an extreme case where there is no electrical power input. Both sides of the equation are equal to zero in this case, suggesting that no other constant term is needed, and that Eq. (1) may be enough for data fitting.

Going back to the specific rate of radiation, convection, and conduction, we have
$$Q_{rad} = \sigma\varepsilon(T_h^4 - T_c^4)A_{eff} \tag{2}$$
$$Q_c = \frac{Q_{fm}Q_{cont}}{Q_{fm} + Q_{cont}} \tag{3}$$

Note that $A_{eff}$ is used instead of $A$ when we need to take the lateral side of the heater-sample complex into account. $A_{eff}$ could be larger than $A$ because the testing film also covers most of heater's lateral side. The lateral side may not be additive to the front and back sides of the heater regarding heat dissipation effectiveness, so it is ignored first while calculating the surface area of the heater (i.e., $A$) but later fitted in the model.

In Eq. (3), the free-molecular conduction (with convection ignored due to the low pressure) is
$$Q_{fm} = \frac{1}{8}\frac{\gamma+1}{\gamma-1}\frac{\bar{v}}{\bar{T}}A_{eff}\alpha p\Delta T \tag{4}$$

And the continuum conduction and convection are concluded in:
$$Q_{cont} = Nu \cdot r\pi\kappa\Delta T \tag{5}$$
Remind that for diatomic gases $\gamma = C_P/(C_P - R)$. Some supplementary formulae for Eq. (4) and (5) are shown below:

$$\bar{v} = \sqrt{\frac{8R_{air}\bar{T}}{\pi}} \tag{6}$$

$$Nu = 0.417Ra^{0.25} + \frac{8}{\pi} \tag{7}$$

$$Ra = \frac{g}{\bar{T}}\frac{\rho(2r)^3}{\mu D}\Delta T \tag{8}$$

$$\rho = \frac{pR_{air}}{\bar{T}} \tag{9}$$

$$D = \frac{\kappa}{\rho C_P} \tag{10}$$

$$\mu = 1.716 \times 10^{-5}\left(\frac{\bar{T}}{273}\right)^{\frac{2}{3}} \tag{11}$$

$$\kappa = 2.38 \times 10^{-4}\,\bar{T}^{0.8218} \tag{12}$$

$$C_P = (28.11 + 1.956 \times 10^{-3}\bar{T} + 4.802 \times 10^{-5}\bar{T}^2 - 1.966 \times 10^{-9}\bar{T}^3) \times \frac{1000}{28.97} \tag{13}$$

From Eq. (2)-(5), we find that $A_{eff}$ and $\Delta T$ show up multiple times. To simplify the fitting, we divide $A_{eff}\Delta T$ on both sides of Eq. (1) and it follows that

$$\frac{U_{heater}I}{A_{eff}\Delta T} = \frac{Q_{rad}}{A_{eff}\Delta T} + \frac{Q_c}{A_{eff}\Delta T} + C' \tag{14}$$

Note that the radiation term on the right side of Eq. (14) is equivalent to $2\sigma\varepsilon(T_h^2 + T_c^2)\bar{T}$, while the second and third terms can also cancel $\Delta T$ using Eq. (4) and (5). In addition, $C'$ in Eq. (14) is not the same as those in Eq. (1), but they are simply related by $C' = C/A_{eff}$. Remind that we have a third fitting parameter, $\alpha$, which is the ultimate goal of the measurement. Meanwhile, $C$ and $A_{eff}$ are dependent on the type of heater, thermal and mechanical adhesion layer, coverage of the sample disk on the heater. Their values are thus lacking consistency and importance for data processing among multiple tests.

In constant-temperature measurements, we maintained the heater-film structure at a certain temperature (usually 40~70 degrees Celsius) and recorded the electrical power's change versus the chamber pressure. Note that the silver paste between the heater and the testing film ensured the two surfaces had few temperature differences. The heater had a linear relationship between its electrical resistance and its surface temperature, as pre-calibrated by a high-resolution thermal camera (TESTO).

At a constant temperature, the radiation heat loss is also a constant, so only all-regime convection/conduction and another constant term resulting from Joule's heating or heat leakage on the sandwich's lateral sides need to be included in the equation. Hence, Eq. (1) can be written as:

$$\frac{Q_c}{\delta A \Delta T} + C - \frac{U_{heater} I}{\delta A \Delta T} = 0 \qquad (16)$$

It is emphasized that $\Delta T$ is a constant term here so it is divided just to make Eq. (16) similar to Eq. (15). A MATLAB trick is used here to express $A_{eff} = \delta A$, which avoids using a very small value (i.e., $A_{eff}$) in the data processing and reduces the fitting error. Recall that $\alpha$ is also a fitting parameter existing in $Q_c$, so we plotted $U_{heater} I / \delta A \Delta T$ versus $p$ and used the nonlinear least squares solver in MATLAB to derive the optimal set of $(\alpha, \delta, C)$.

Briefly, a thin disk-shaped heater was hung in the center of a steel vacuum chamber as shown in Fig. S5, where the chamber pressure could be adjusted continuously from $10^{-4}$ to 200 Pa (i.e., $7.5 \times 10^{-7}$ to 1.5 Torr). During every set of measurements, we gradually increased the chamber's pressure and the electrical energy input while keeping the heater's temperature constant.

An example of raw data and its processing on the platinum surface is shown in Fig. S6. The temperature of the heater-film was well-controlled at around 400.9 K with a coefficient of variation (CV) of 0.14%. The fitting result gives values of $(\alpha, \delta, C)$ as (0.75, 1.22, 4.09) and 95% confidence intervals as (0.68~0.82, 1.20~1.23, 3.99~4.19). We conducted at least two experiments for every material to ensure reproducibility. A summary of all TAC values and their 95% confidence intervals are listed in Table S2.

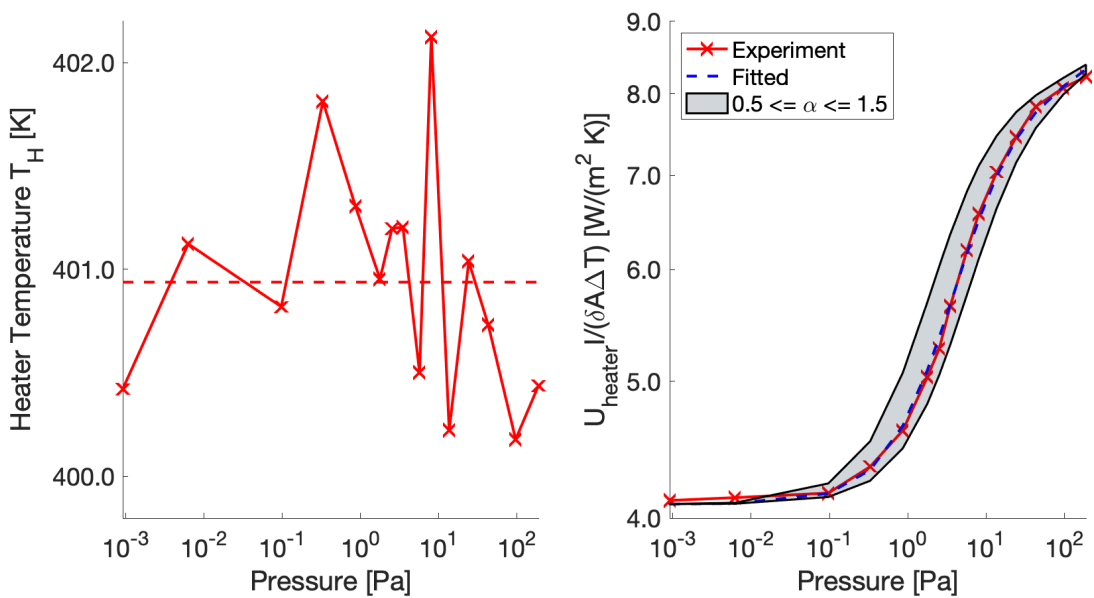

FIG. S6. Results of a constant-temperature TAC measurement on the mylar-platinum surface. (a) The temperature control. (b) Effective conductivity of heater-film system and the fitted curve. Note that the temperature uncertainty across the heater is not shown in (a), and that the grey shaded area in (b) represents where curves should fall with $0.5 \leq \alpha \leq 1.5$.

TABLE S2. Results of TAC values (bold) and confidence intervals (in brackets).

| MATERIAL | TEST 1 | TEST 2 | TEST 3 OR LITERATURES |
|---|---|---|---|
| Gold | **1.14** (1.04~1.23) | **1.01** (0.93~1.13) | ~1 [4] |
| Platinum | **0.75** (0.68~0.82) | **0.69** (0.64~0.76) | ~0.75 [2] |
| Mylar | **0.80** (0.75~0.85) | **0.76** (0.71~0.82) | **0.75** (0.71~0.81) |
| Alumina | **0.82** (0.74~0.90) | **0.77** (0.71~0.86) | **0.80** (0.74~0.85) |
| CNT | **1.05** (0.96~1.15) | **1.13** (1.05~1.22) | **1.03** (0.95~1.15) |
| Germanium | **0.90** (0.83~0.97) | **0.93** (0.85~1.01) | **0.88** (0.81~0.96) |
| Germanium-CNT | **0.99** (0.93~1.06) | **0.96** (0.88~1.05) | **0.95** (0.89~1.01) |

Therefore, in the following theoretical model of the light-flyer, we set the TAC for the top (i.e., alumina) and bottom (i.e., germanium) as 0.8 and 0.9, respectively.

*6. Theoretical modeling for light-flyers*

The photophoretic force of a disk with different thermal accommodation coefficients ($\alpha$) on its top and bottom is called $\Delta\alpha$ force. According to Rohatschek [5] and Azadi's theories [6], we built up the light-flyer disk's model by considering a two-dimensional disk's heat dissipation and different force mechanisms in the free-molecular and continuum regimes.

TABLE S3. Descriptions of symbols in light-flyer disk modeling. All symbols, unless dimensionless or specified, are in SI units.

| SYMBOL | DEFINITION |
|---|---|
| $T_c$ | Environmental temperature |
| $T_h$ | Temperature of the light-flyer disk |
| $\bar{T}$ | Average temperature, i.e., $(T_h + T_c)/2$ |
| $\Delta T$ | Temperature difference, i.e., $(T_h - T_c)$ |
| $\beta$ | Thermal expansion coefficient for ideal gas, i.e., $1/\bar{T}$ |
| $r$ | Radius of the disk |
| $\alpha_{top}, \alpha_{bot}$ | Top (bottom) surface thermal accommodation coefficient |
| $\Delta\alpha$ | Difference in thermal accommodation coefficient, i.e., $\alpha_{bot} - \alpha_{top}$ |
| $\varepsilon$ | Uniform surface emissivity |
| $\epsilon_{vis}$ | Optical absorptivity of the disk |
| $A$ | Bottom area of the disk, i.e., $\pi r^2$ |
| $p$ | Ambient pressure |
| $R_{air}$ | Ideal gas constant of air, i.e., 287.05 J/ (kg K) |
| $\bar{v}$ | Average molecular velocity of air |
| $\mu$ | Dynamic viscosity of air |
| $\rho$ | Density of air |
| $D$ | Thermal diffusivity of air |
| $\gamma$ | Adiabatic index of air (diatomic gases) |
| $\kappa$ | Thermal conductivity of air |
| $Ra$ | Rayleigh number |
| $Nu$ | Nusselt number |
| $\sigma$ | Stefan-Boltzmann constant, i.e., $5.67 \times 10^{-8}$ W/ (m² K⁴) |
| $I$ | Incident light irradiance |
| $Q_{rad}$ | Radiation loss |
| $Q_c$ | Convection and conduction loss in all regimes |
| $F$ | Photophoretic force in all regimes |
| $F_{fm}$ | Photophoretic force in the free-molecular regime |
| $F_{cont}$ | Photophoretic force in the continuum regime |
| $\kappa_s$ | Thermal slip coefficient, i.e., 1.14 [ref] |
| $C_{fm}, C_{cont}$ | Dimensionless coefficients for photophoretic forces |
| $h$ | Altitude (in the unit of kilometer) |

The light-flyer disk's temperature needs to be solved first for the following force derivation. Similar to the Eq. (1), the energy equilibrium of the disk can be described as:

$$A\epsilon_{vis}I = Q_{rad} + Q_c \tag{17}$$

Where $Q_{rad} = 2A\sigma\varepsilon(T_h^4 - T_c^4)$ and $Q_c$ can be calculated as shown in Eq. (3) – (13) by substituting all $A_{eff}$ with $2A$.

As mentioned in the main text, the photophoretic force at all pressures can be described as a combination of those in free-molecular and continuum regimes:

$$\frac{1}{F} = \frac{1}{C_{fm}F_{fm}} + \frac{1}{C_{cont}F_{cont}} \tag{18}$$

In the free-molecular regime, the photophoretic force is proportional to the difference in TAC and can be derived as:

$$F_{fm} = \frac{A}{4T_c}p\Delta\alpha\Delta T \tag{19}$$

In the continuum regime, the photophoretic force is more complicated:

$$F_{cont} = 16\mu \cdot 6\kappa_s T_h \frac{\kappa}{r\rho p\bar{v}} \frac{\gamma-1}{\gamma+1} \frac{\Delta\alpha}{\bar{\alpha}} \frac{\Delta T}{T_c} \tag{20}$$

The temperature of the light-flyer disk, photophoretic force, and resulting light irradiance are solved sequentially and iteratively in Python.

According to the temperature profile in the mesosphere [7][8], a 15-order polynomial is used to fit the inverse temperature as a function of the altitude (in the unit of kilometer):

$$T_c^{-1}(h) = \begin{bmatrix} -4.59 \times 10^{-29} \\ 4.02 \times 10^{-27} \\ 1.49 \times 10^{-23} \\ -7.94 \times 10^{-21} \\ 2.02 \times 10^{-18} \\ -3.15 \times 10^{-16} \\ 3.27 \times 10^{-14} \\ -2.33 \times 10^{-12} \\ 1.15 \times 10^{-10} \\ -3.86 \times 10^{-9} \\ 8.53 \times 10^{-8} \\ -1.15 \times 10^{-6} \\ 8.15 \times 10^{-6} \\ -2.28 \times 10^{-5} \\ 9.91 \times 10^{-5} \\ 3.47 \times 10^{-3} \end{bmatrix}^T \begin{bmatrix} h^{15} \\ h^{14} \\ h^{13} \\ h^{12} \\ h^{11} \\ h^{10} \\ h^9 \\ h^8 \\ h^7 \\ h^6 \\ h^5 \\ h^4 \\ h^3 \\ h^2 \\ h \\ 1 \end{bmatrix} \tag{21}$$

On the other hand, the pressure as a function of the altitude can be calculated from:

$$p(h) = 101325 e^{-\frac{g}{R_{air}}\zeta(h)} \tag{22}$$

Where $\zeta(h) = \int T_c^{-1}(h)dh$.

Using the aforementioned theory, the operational pressure range and surface temperature of germanium-coated disks with diameters of 3 and 4 cm are shown in Fig. S7.

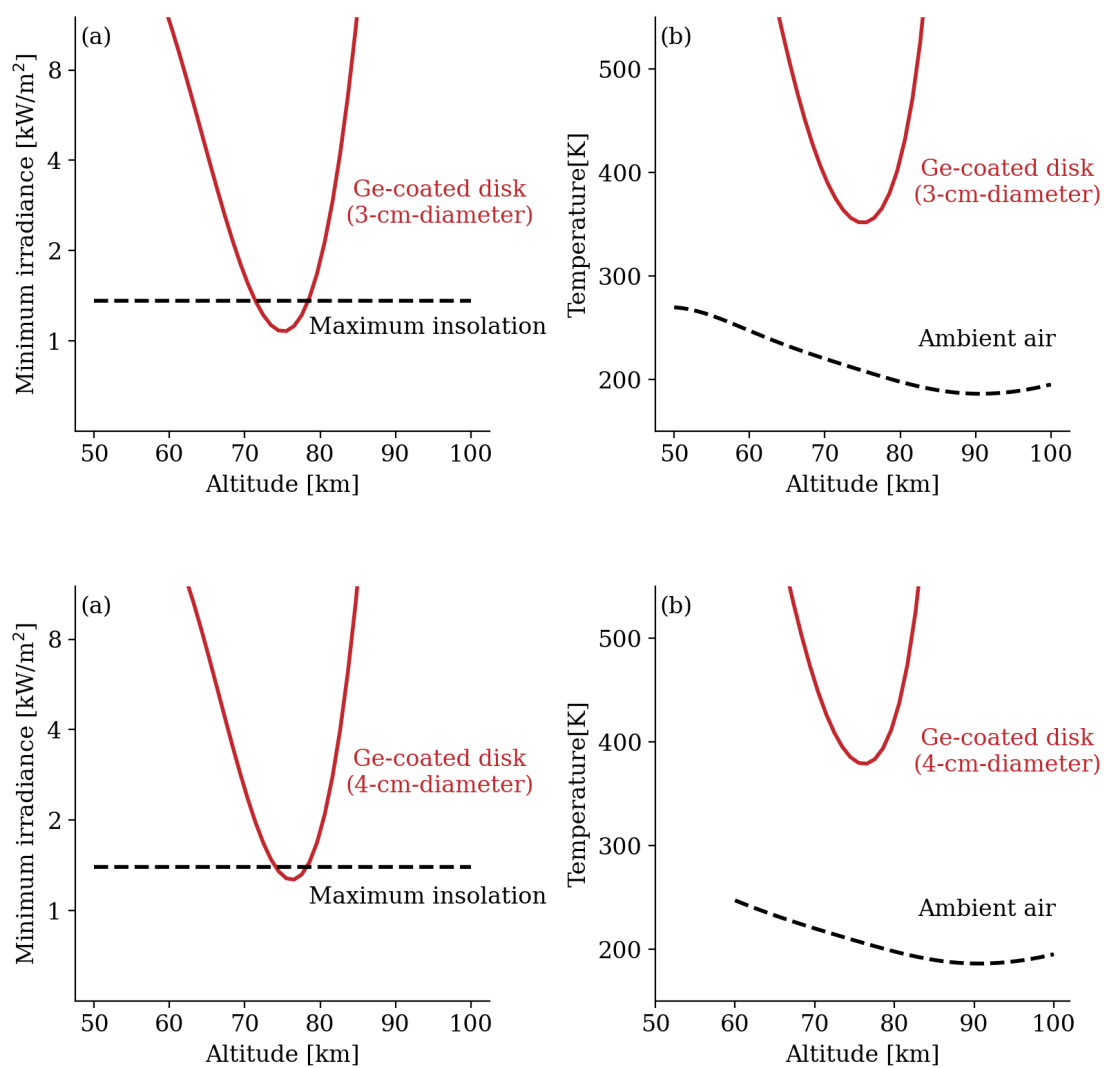

FIG. S7. Same as Fig. 5 for germanium-coated disks with diameters of (a-b) 3 and (c-d) 4 cm.

## 7. Predictions of light-flyers for mesospheric applications

The operational altitude ranges and payload capabilities of light-flyers depend on their size and are crucial for future atmospheric sensing and data communication. In this context, we define *operational altitude* as the range of altitudes at which a light-flyer can levitate, *optimal altitude* as the altitude at which a light-flyer can carry the maximum payload, and *relative payload* as the ratio of the payload a light-flyer can carry to its own weight. According to Fig. S8, it is evident that at the maximum possible insolation (i.e., 1.36 kW/m$^2$), light-flyers smaller than 4.75 cm in diameter can be levitated with positive payloads within their operational altitude ranges. Notably, light-flyer disks smaller than 0.5 cm in diameter can carry payloads heavier than their own weights. Though small in individual capacity, a network of light-flyers can be combined to carry heavier payloads by dividing it into several parts. The light-flyer network envisions a functionalized system of aerial vehicles, antennas, and microelectromechanical sensors, where a 3-cm-diameter germanium-coated disk is the reference design due to its largest payload capacity of 0.21 mg and medium-large relative payload of 0.30.

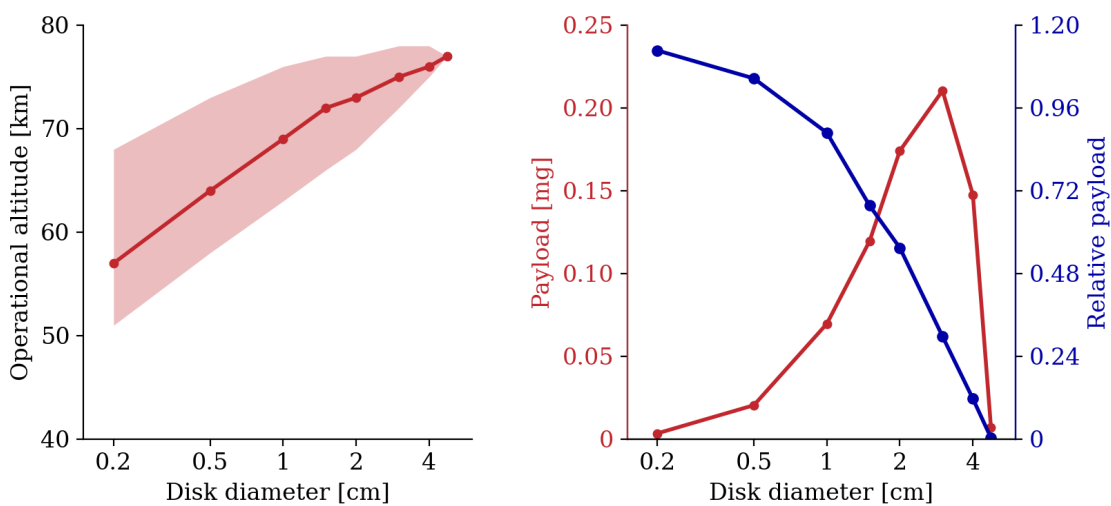

FIG. S8. Predictions for mesospheric applications. Model parameters are the same as Fig. 5. (a) Plots of operational altitudes for various sizes of light-flyer disks. The dot marker denotes the optimal altitude, and the shaded area represents the whole operational altitude range. (b) Plots of payloads (left axis) and relative payloads (right axis) for various sizes of light-flyer disks.